\documentclass[manuscript]{aastex}

\newcommand{\etal}{et al. }
\newcommand{\be}{\begin{equation}}
\newcommand{\ee}{\end{equation}}

\newcommand{\aeta}[3]{ 19#1, A\&A,  #2, #3}

\newcommand{\aspj}[3]{ 19#1, ApJ, #2, #3}

\newcommand{\sph}[3]{ 19#1, Solar Phys., #2, #3}

\begin{document}

\title{The Counter-kink Rotation of a Non-Hale Active Region}

\author{M.C. L\'opez Fuentes \altaffilmark{1},}
\affil{Instituto de Astronom\'\i a y F\'\i sica del Espacio, IAFE,
CC. 67 Suc. 28, 1428 Buenos Aires, Argentina,}

\author{P. D\'emoulin,}
\affil{Observatoire de Paris, section Meudon, DASOP, URA 2080 (CNRS),
F-92195 Meudon Principal Cedex, France,}

\author{C.H. Mandrini \altaffilmark{2},}
\affil{Instituto de Astronom\'\i a y F\'\i sica del Espacio, IAFE,
CC. 67 Suc. 28, 1428 Buenos Aires, Argentina,}

\and

\author{L. van Driel-Gesztelyi}
\affil{Konkoly Observatory, H-1525, Budapest, Hungary}

\altaffiltext{1}{Fellow of the CONICET, Argentina}
\altaffiltext{2}{Member of the Carrera del Investigador Cient\'\i fico,
CONICET, Argentina}

\authoremail{lopezf@iafe.uba.ar}

\begin{abstract}
  We describe the long-term evolution of a bipolar non-Hale active region which was observed from October, 1995, to January, 1996. Along these four solar rotations the sunspots and subsequent flux concentrations, during the decay phase of the region, were observed to move in such a way that by December their orientation conformed to the Hale-Nicholson polarity law.  The sigmoidal shape of the observed soft X-ray coronal loops allows us to determine the sense of the twist in the magnetic
configuration.  This sense is confirmed by extrapolating the observed photospheric magnetic field, using a linear force-free approach, and comparing the shape of computed field lines to the observed coronal loops.  This sense of twist agrees with that of the dominant helicity in the solar hemisphere where the region lies, as well as with the evolution observed in the longitudinal magnetogram during the first rotation. At first sight the relative motions of the spots may be
miss-interpreted as the rising of an $\Omega$-loop deformed by a kink-instability, but we deduce from the sense of their relative displacements a handedness for the flux-tube axis (writhe) which is opposite to that of the twist in the coronal loops and, therefore, to what is expected for a kink-unstable flux-tube.  After excluding the kink instability, we interpret our observations in terms of a magnetic flux-tube deformed by external motions while rising through the convective zone.  We compare our results with those of other related studies and we discuss, in particular, whether the kink instability is relevant to explain the peculiar evolution of some active regions.

\end{abstract}


\section{Introduction}
\label{introduction}

   Solar active regions (ARs), as observed at the photospheric level,
consist of large areas with a strong magnetic field concentration
($\geq$ 1000 G) as compared to the quiet-region average field.  Since,
in general, an AR is formed by two of such areas of opposite magnetic
polarity, it has long been believed that they are the manifestation of
the emergence of a flux tube formed from the toroidal magnetic field
originated at the base of the convective zone (e.g.  Parker 1993,
Weiss 1994).  The typical shape of this flux tube is that of the
letter $\Omega$, which results from the rising of a buoyant magnetic
flux-tube (e.g.  Zwaan 1987 and references therein).  As the flux tube
crosses the photospheric surface, two flux concentrations of opposite
polarity appear and progressively diverge from each other in an
approximate East-West direction.  Such bipolar ARs will, in general,
obey the Hale-Nicholson polarity law (see e.g. Zirin 1988).

  However, there are several observational examples of ARs, or emerging
bipoles within ARs, disobeying the Hale-Nicholson's law (e.g.  Tanaka
1991, Lites et al.  1995, Leka et al.  1996, Pevtsov \& Longcope
1998).  The evolution of such photospheric flux concentrations has
been explained in terms of the rising of very distorted flux-tubes.
Tanaka (1991) proposed a model implying the emergence of a ``knotted''
flux-tube to explain the observed evolution of two very active
$\delta$ configurations.  Though not explicitly mentioned in his
paper, the shape of the tube resembled that of a kink-unstable flux
tube (see Linton et al.  1998, 1999, and Fan et al.  1998, 1999, for
theoretical developments).  Van Driel-Gesztelyi and Leka (1994) and
Leka et al.  (1996) analyzed series of magnetograms corresponding to
an AR, where strong flux emergence was observed, and concluded that
the proper motions for several emerging bipoles were consistent with
the rising of kink-deformed flux tubes.  More recently, Pevtsov \&
Longcope (1998) interpreted the magnetic data and soft X-ray images of
a pair of ARs, observed during two solar rotations, as an evidence of
the emergence of a single magnetic system resembling a kinked
flux-tube.  In the case of Lites et al.  (1995), the evolution of a
$\delta$ configuration was explained as resulting from the ascension
of a nearly closed system of twisted magnetic field, unrelated to the
kink instability.  More generally, Weart (1970, 1972) noticed the
almost random distribution of the starting tilt of emerging bipoles,
which subsequently became more parallel to the equator, and proposed
that this was caused by the emergence of twisted flux-tubes.

  We describe here the magnetic field evolution of a region formed by a pair
of sunspots, which, at its appearance on the disk, disobeyed the
Hale-Nicholson polarity law for solar cycle 22.  For this analysis we
used a set of magnetograms obtained at Kitt Peak National Solar
Observatory (KPNO), and soft X-ray images obtained with the Soft X-ray
Telescope (SXT) on board the {\it Yohkoh} satellite.  We followed the
region along four solar rotations, from its appearance on the solar
disk in October 1995, until its decay and disappearance in January
1996.  Along these three months, we observed that the following spot
rotated relative to the preceding one in such way that by the end of
this period both flux concentrations were aligned according to the
Hale-Nicholson law.  In \S \ref{evolution} we describe the temporal
evolution of the studied region, while in \S \ref{interpretation} we
discuss a possible model to explain such an evolution.  Finally, in \S
\ref{conclusion} we summarize our conclusions and we discuss the role
of the kink instability in the peculiar behavior of some active
regions.

\section{Non-classical evolution of the active region}
\label{evolution}

\subsection{Description of the data}
\label{descrip_data}

   A set of 23 line-of-sight magnetograms obtained at KPNO has been used to
follow the evolution of the non-Hale region NOAA 7912.  The
magnetograph of KPNO (Livingston et al.  1976) provides daily full
disk longitudinal magnetic field maps with a spatial resolution of
around 1".  The maps used in the present study correspond to the
months of October, November and December 1995, and January 1996.  We
chose approximately 6 magnetograms per rotation, when available,
around the day of the central meridian passage (CMP) of the region of
interest.

  We complemented the magnetic data set with soft X-ray full disk
images obtained with the Yohkoh/SXT (Tsuneta et al.  1991).  We chose
these images at times close to the KPNO magnetograms, and we coaligned
these two data sets in order to follow changes in the coronal magnetic
field structure as the region evolved.

\subsection{Long-term evolution of the region}

\begin{figure*}[t]
\centering
\hspace{0.cm}
\includegraphics[bb=100 50 475 785,width=8.5cm]{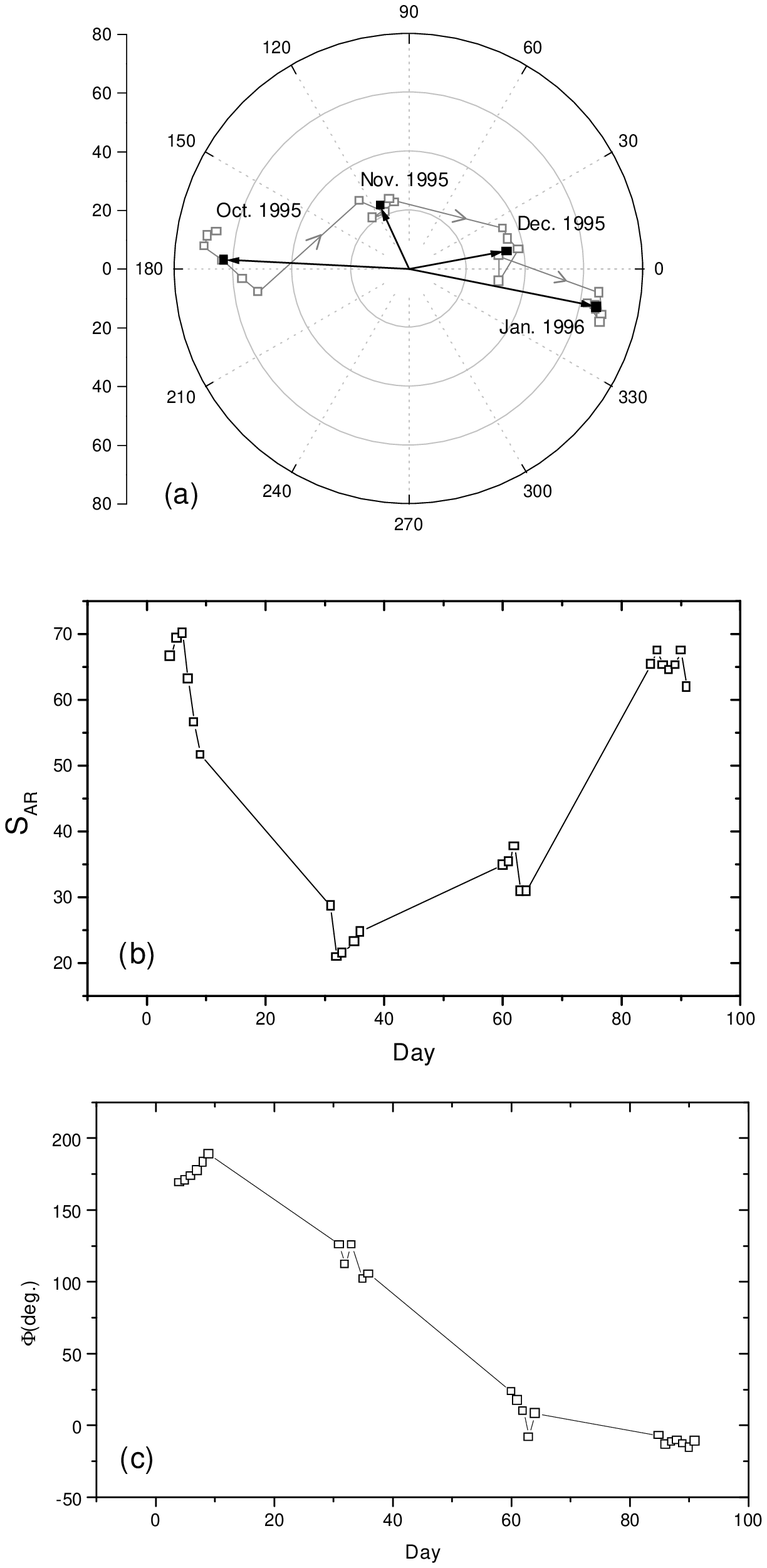}
       \caption{Relative evolution of the leading polarity with respect to the
following one (only the magnetograms closer than 33 deg. from
CMP have been used in order to avoid important projection
corrections). (a) Polar plot (see text for an explanation).
The arrows on the light grey lines show the way the time proceeds.
(b,c) Time evolution of the size $S_{AR}$ of the active region and
of the angle $\Phi_{AR}$ as defined in the text.
The vertical axes in (a) and (b) are expressed in
Mm. Day number one corresponds to October 10, 1995 in (b) and (c).}
\label{f-rot}
\end{figure*}

  A bipolar sunspot group, NOAA 7912,  was observed on the solar disk on
October 10, 1995, at S10 E76.  This AR had non-Hale polarity relative
to the Hale-Nicholson polarity law for solar cycle 22.  During its
disk transit the negative polarity (located westward) appeared
concentrated, while the positive one was more diffuse (see Fig.
\ref{f-evolution} right panel).  CMP occurred on October 15, 1995.  In
the following solar rotation, at the same latitude and corresponding
longitude according to the solar rotation rate, another non-Hale
bipolar sunspot group was observed.  This group was numbered as NOAA
7921, and we show in the next paragraph that it corresponds to AR 7912
of the previous rotation.  During this second rotation the positive
and negative polarities appeared to be closer together (see e.g.  Fig.
\ref{f-rot}.a and b).  Another particular feature of this AR 7921 is
that the trailing (positive) spot lies closer to the solar equator
than the leader (negative) spot, contrary to what is expected from
Joy's law (see e.g.  Zirin 1988).  During the third solar rotation
(December 1995), we observe a bipolar AR (NOAA 7930) traversing the
solar disk at the same latitude and corresponding longitude as AR
7912.  In this case the leading and trailing spots obey the
Hale-Nicholson polarity law, but the leading (positive) magnetic field
concentration appears much more dispersed than the trailing (negative)
one.  From the coincidence in location (see also next paragraph), we
conclude that this is still the remnant of AR 7912 which we now
observe in its decaying phase.  In the fourth rotation (January 1996)
no AR was identified at that position, though we observed a negative
and a positive flux concentration oriented almost parallel to the
solar equator with the leading positive field more dispersed than the
trailing negative one.

\begin{figure*}[t]
\centering
\hspace{0.cm}
\includegraphics[bb=30 100 560 770,width=12.cm]{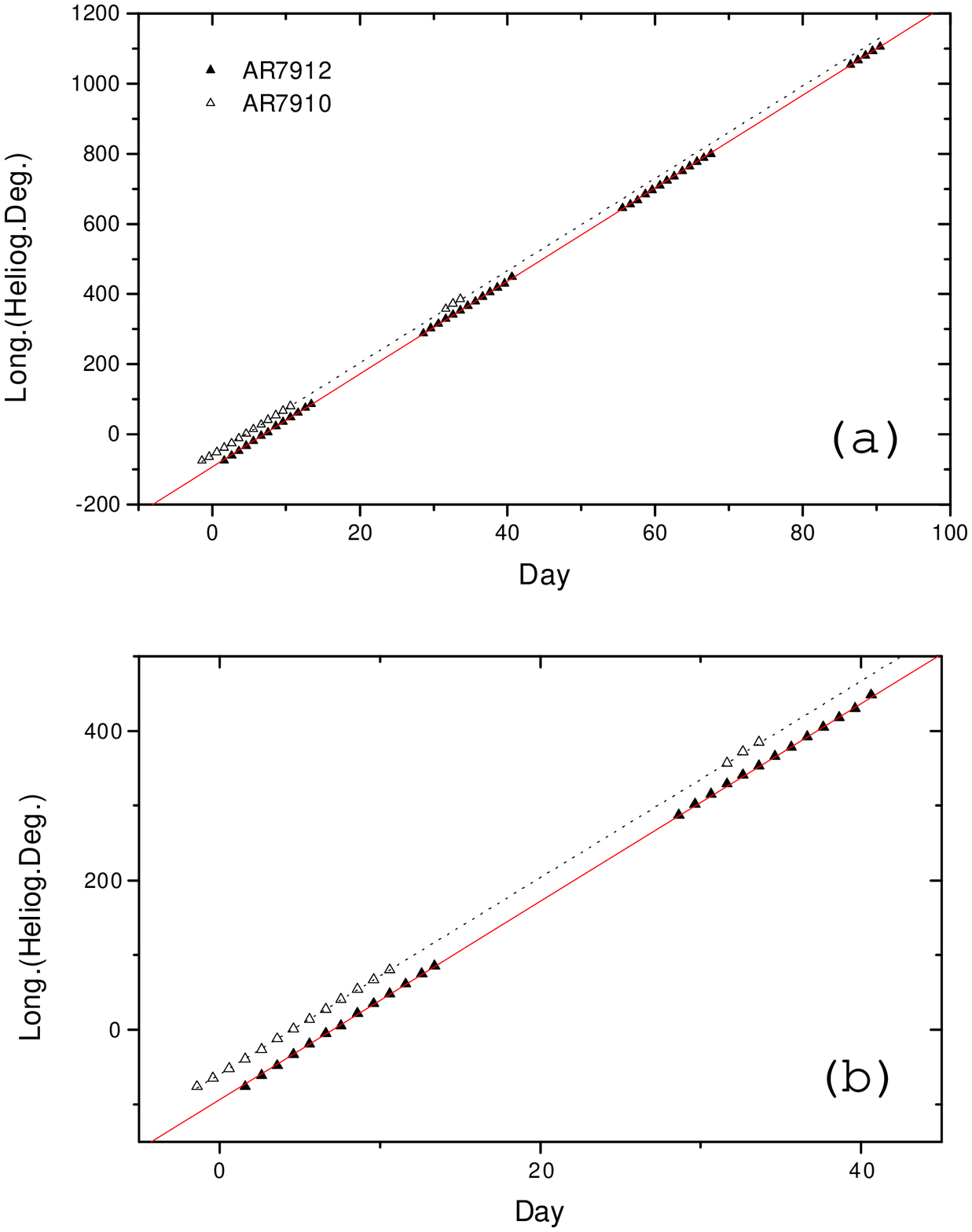}
       \caption{Synodic longitudes for the ARs 7910 and 7912 and their subsequent appearances
on the solar disk, as a function of time (only few days around the CMP are
plotted in the case when no AR was identify at that location).
Day number one corresponds to October 10, 1995. A least square fit with a
constant rotation speed is added. (a) four rotations, (b) zoom on the first
two rotations.}
\label{rot_rate}
\end{figure*}

   We now argue, following similar arguments as Pevtsov and Longcope
(1998), that AR~7921, AR~7930 and the subsequent positive and negative
flux concentrations observed during January are the recurrences of
AR~7912.  In Fig.  \ref{rot_rate} we show the successive synodic
longitudes for the three ARs, taken from the Solar Geophysical Data,
and those for the bipolar flux concentrations observed during January,
as measured using KPNO magnetograms.  It is clear that all the points
lie on a straight line, whose slope gives the value of the synodic
solar rotation rate ($\omega$) for the corresponding latitude
($\approx$ 10 deg).  Thus, we conclude that during the four rotations
the renamed ARs and flux concentrations were located at the longitude
and latitude where the recurrent remnants of AR 7912 were expected to
be.  In our case $\omega$ turns out to be 13.25 $\pm$ 0.02 deg/day.
Howard (1990), using Mt.  Wilson magnetograms in the period 1967-1988,
found for the latitude of 10$^{o}$ a synodic rotation rate 13.015
$\pm$ 0.038 for all ARs and 13.05 $\pm$ 0.15 for reversed polarity
groups.  The rotation rate we determined is around the upper limit for
the latter.  It is noteworthy that reversed polarity groups represent
about 10 \% of all ARs, and there is no evidence for a latitudinal
dependence of their rotation rate (Howard, 1990).

   Looking at the position of the AR closest to AR~7912, we find that
during the first rotation this was AR 7910, located at the same
latitude (S10) and 33 degrees to the West of AR 7912; this was a Hale
region.  Considering the errors bars, it appears very unlikely that we
could confuse AR 7921 with the reappearance of AR 7910 (Fig.
\ref{rot_rate}), even more taking into account that AR 7921 is still a
non-Hale region (Fig.  \ref{f-evolution}, right panel).  During the
third rotation (in December) AR 7930 was practically the only active
region in the southern hemisphere.

   The lifetime of a sunspot both from an observational point of view
(Petrovay \& van Driel-Gesztelyi 1997), and according to the sunspot
turbulent erosion model (Petrovay \& Moreno-Insertis 1997), is around
41 days in the case it has a maximum area of 410 MSH (millionth of
solar hemisphere), which approximately corresponds to the maximum area
of the negative spot including its umbra and penumbra.  Thus, the
negative sunspot observed during the first rotation (AR 7912) is
likely to have survived till the second rotation (AR 7921); then its
magnetic flux progressively dispersed during the next rotations.  This
supports that the negative polarity observed in the four rotations is
indeed formed by the same magnetic flux.  Moreover, the lifetime of an
isolated AR can be as long as 7 months; after that its dispersed
magnetic field becomes indistinguishable from the background field
(see the review by van Driel-Gesztelyi 1998).  Therefore, we find it
unlikely that AR 7912 had decayed and disappeared on the invisible
side of the Sun, being then replaced by the emergence of another
magnetic flux tube during the three consecutive rotations.

   From the stringent coincidences in location, the recurrence
and expected life span of active regions and sunspots of similar size,
we conclude that, along these three months, we have been observing the
growth, maturity and decay of the same AR, which we will call from now
AR 7912.  Its evolution is exemplified in Fig.  \ref{f-evolution}
(right panel) in which we show one magnetic map per rotation at CMP.
The most striking features of this figure are the way in which the
positive polarity seems to rotate around the negative one, and the
variable mean distance between the flux concentrations.

 We extrapolated the observed photospheric longitudinal magnetic field
using a linear force-free approach ($\vec \nabla \times \vec B =
\alpha \vec B$, see e.g.  D\'emoulin et al.  1997), where $\alpha $ is
determined by the best fit between the soft X-ray loops and the
computed field lines.  We computed in the local orthogonal frame, that
is ($x,y$) parallel to the photosphere and $z$ perpendicular to it,
the dipolar size of the AR ($S_{AR}$) and the angle ($\Phi_{AR}$)
formed by the line joining the mean position of the positive and
negative concentrations (see Eq.(\ref{Pos})) with the local parallel.
We define the dipolar size of the region as the flux-weighted mean
distance between opposite polarity fields which are stronger than
some limit $B_{min}$,
\be
 S_{AR} = \sqrt{(X_p - X_n)^2 + (Y_p - Y_n)^2} \, ,
\ee
and $\Phi_{AR}$ as
\be
 \Phi_{AR} = \arctan{(Y_p - Y_n)/(X_p - X_n)} \, ,
\ee
where $X_p$ and $Y_p$ ($X_n$ and $Y_n$) give the mean position of the
positive (negative) concentration,
\begin{eqnarray}\label{Pos}
  X_p = {\sum_{B_z > B_{min}} x B_z \over \sum_{B_z > B_{min}} B_z} \, ,
  Y_p = {\sum_{B_z > B_{min}} y B_z \over \sum_{B_z > B_{min}} B_z} \,
\end{eqnarray}

  We computed $S_{AR}$ and $\Phi_{AR}$ for the 23 line of sight
magnetograms included in our study.
To be sure that the trend followed by $S_{AR}$ and $\Phi_{AR}$ is not
affected by the value of $B_{min}$ considered in the computations, we
have taken field strengths $| B_z | > 10$ G, $50$ G and $100$ G.  To
check the absence of any systematic bias introduced by the magnetic
extrapolation, we repeated the analysis replacing $B_z $ by the
observed longitudinal field.  We found the same kind of temporal
variation in all cases.  Fig.  \ref{f-rot}.a shows a polar diagram,
for $| B_z | > 100$ G, where the displacement of the positive polarity
around the negative one is clearly seen along the four solar
rotations.  The angle $\Phi_{AR}$ is measured from West to East
(counterclockwise), the center of the polar plot corresponds to the
mean position of the negative polarity, while the light grey squares
give the successive locations of the positive polarity and the black
squares represent the average $\Phi_{AR}$ angle for each solar
rotation.  Moreover, the distance between the light grey squares and
the center of the polar plot corresponds to $S_{AR}$, while the long
arrows joining the black squares to the centre correspond to the
average $S_{AR}$ for each rotation.  We show the time evolution of
$S_{AR}$ and $\Phi_{AR}$ in a more explicit way in Figs.
\ref{f-rot}.b and c, respectively.
The rotation by half a turn during four solar rotations as well as
the approach of two opposite polarities during the first two rotations,
followed by the increase in their separation during the third and
fourth rotations, are abnormal characteristics of the AR 7912.
From this analysis we deduce below the possible origin of this
peculiar evolution.

\section{Interpretation in terms of a rising flux-tube}
 \label{interpretation}

\subsection{Emergence of the flux tube}

   The photospheric magnetic evolution of the region allows us to deduce the
shape of the corresponding emerging flux-tube, assuming that all four
ARs were formed by a single $\Omega$-loop.  Unfortunately, we have
neither a way to estimate the velocity of the emergence nor its change
with time, so the conclusions drawn below concerning the shape of the
flux tube may be affected by an arbitrary factor implying an extension
(or compression) in the vertical direction.  This has, however, no
influence on the handedness of the tube axis and, therefore, on our
conclusions.

  The evolution of the photospheric magnetic field during the four
solar rotations is not compatible with the emergence of a simple
planar $\Omega$-loop, but it has to be interpreted as the emergence of
a magnetic flux-tube which is deformed as shown in the left panel of
Fig.  \ref{f-evolution}.  The flux-tube is drawn as it was when in the
convection zone.  As it progressivelly emerges, the
intersections of both of its feet with the photospheric level
(positive and negative polarities) rotate as observed in the
magnetograms (Fig.  \ref{f-evolution}, right panel).  The portion
above the photosphere will relax to an almost force-free field state
and expand to fill the available volume; this latter evolution in not
shown in Fig.  \ref{f-evolution}.

  Provided that we have no information on the velocity of emergence,
the observed flux tube evolution may result from one of the two following
scenarios.  In the first one, only the top part of the flux tube is
buoyant enough to emerge above the photosphere.  When the upward
evolution stops, the further observed rotation may result from the
magnetic tension of the flux tube (such force tends to bring the flux
tube back to the planar configuration defined by the flux-tube feet
rooted deeply in the convective zone).  A second scenario is that, in
the upper part of the convective zone, the buoyancy of the flux-tube
feet is still present (while lower than at its summit).  As a result
of this, the upward speed of the flux-tube decreases with time but
does not vanish (until the flux tube is eroded by turbulence).  Unless
both processes have similar time scales, the continuous rotation of
the positive polarity relative to the negative one (see Fig.
\ref{f-rot}) favor a scenario driven by the process of continuous
emergence, though further data are needed to confirm this.  With such
hypothesis, the magnetic flux-tube is indeed emerging for a much
longer time than the AR formation which, on average, is completed in
about 5 days (Harvey 1993).  The two cases analysed by Pevtsov \&
Longcope (1998) imply indeed flux emergence during two solar
rotations.

  At this point it is noteworthy that the usual expression ``flux emergence''
is, in general, used to refer merely to the increase of the vertical
photospheric magnetic flux (for both polarities).  However, in the
scenario proposed by us, the flux tube keeps emerging for several
months, well beyond the first few days during which the apex of the
flux tube traverses the photosphere.  The continuing upward motion of
the flux tube will not change the magnetic flux in the photosphere
(because $\vec{\nabla } \vec{B}=0$), thus no conventional ``flux
emergence'' will be observed in the magnetic map.  In order to
emphasize this different meaning, we will use the expression
``flux-tube emergence'', rather than ``flux emergence'', to refer to
the full evolution of the flux tube as it is rising through the
photosphere.

\subsection{Magnetic twist}
\begin{figure*}[t]
\centering
\hspace{0.cm}
\includegraphics[bb=55 335 550 590,width=16.cm]{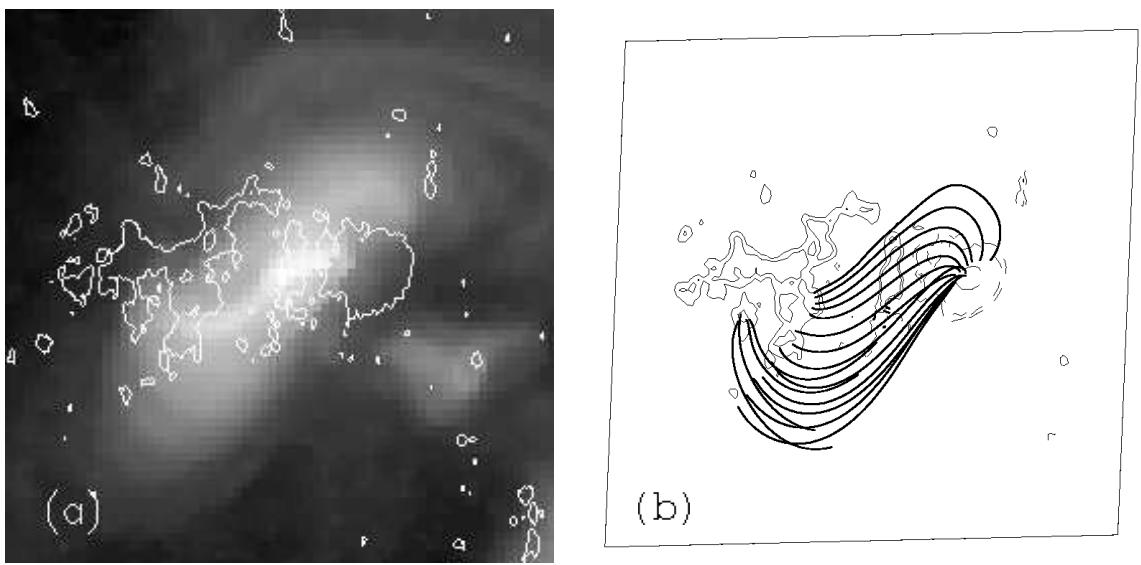}
       \caption{Determination of the twist of the coronal portion of the flux tube.
(a) Section of a soft X-ray full disk image centered at the location of
AR 7912 from Yohkoh/SXT showing an S-shaped
sigmoid. The image was obtained on October 16, 1995, at 15:58 UT.
An isocontour ($\pm 100$ G) of the line of sight magnetic field ($B_l$) has
been added as a reference. (b) Linear force-free magnetic extrapolation,
the best agreement with the SXT observations was found for
$\alpha$ = 0.03 Mm$^{-1}$. The figure is a three dimensional view of the
AR in the observer's perspective. Three isocontour levels of $B_l$
($\pm$ 100, 200, 1000 G) are shown with positive and negative values
drawn with solid and dashed lines, respectively. The size of the region
in both figures is 200 Mm $\times$ 200 Mm. North is up and West is to
the right.}
\label{f-handedness}
\end{figure*}

  Besides its global shape, another important parameter which
characterizes a flux tube is the amount of twist it brings up.  We
have no transverse field measurements available to determine the
direction of the electric currents, but SXT images obtained mainly
during the first three rotations show S-shaped sigmoids at coronal
heights.  This may be an indication of the presence of positive twist
in the magnetic field (see e.g.  Fig.  \ref{f-handedness}.a), although
the loops corresponding to some particular potential (i.e.  without
twist) magnetic configurations can also display sigmoidal shape
(Fletcher et al.  2000).  From our linear force-free extrapolations,
the value of $\alpha$ turns out to $0.03 $ Mm$^{-1}$ (positive sign)
for the data shown in Fig.  \ref{f-handedness}.  Our results show that
the S-shape is really determined by the coronal currents (and not by a
particular distribution of the vertical component of the photospheric
field).  Therefore, AR 7912 follows the hemispherical chirality-rule:
a positive twist is dominant in the southern hemisphere (Seehafer
1990, Pevtsov \etal 1995).

\begin{figure*}[t]
\centering
\hspace{0.cm}
\includegraphics[bb=60 215 525 650,width=15.cm]{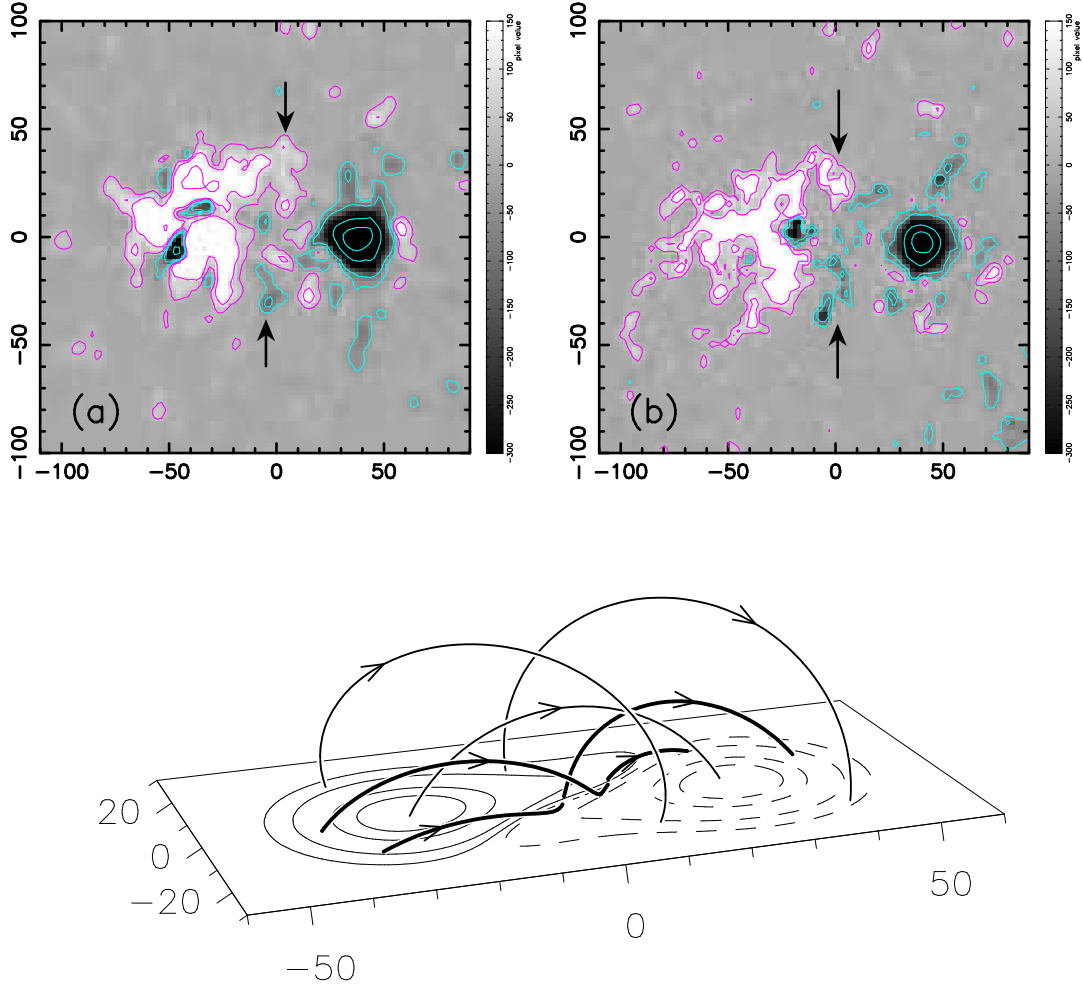}
       \caption{Evolution of the photospheric vertical field during the first
rotation: (a) on October 13, (b) on October 15 (white/black
corresponds to the positive/negative polarities, respectively).
During few days (see Fig.  3) the rotation was in the positive
direction.  This is explained by the contribution of the azimuthal
field component ($B_{\theta}$) to the vertical magnetic component,
during the initial emergence of the flux tube with positive twist as
sketched in (c).  The presence of twist in the $\Omega$-loop implies
the presence of ``tongues'' (pointed with arrows in (a) and (b)) in
the vertical field when the upper part of the tube is emerging at
the photosphere. All axes are in Mm.}
\label{f-pos-rot}
\end{figure*}

   What is the magnetic signature of an emerging twisted flux tube at the
photospheric level?  To simplify the description, we discuss below the
expected evolution for an AR located at disk centre, but this can be
extended to regions in different locations on the solar disk as well.
For the emergence of untwisted $\Omega$-loops, a series of
magnetograms of the vertical field will show the classical appearance
of a bipole, followed by the separation of the two opposite magnetic
polarities.  The magnetograms simply show the evolution of the
vertical component of the magnetic field directed along the tube.
However, when the flux tube is twisted an asymmetry appears in the
magnetogram due to the contribution of the azimuthal component to the
observed vertical component of the field.  The result is schematically
depicted in Fig.  \ref{f-pos-rot}.c for the case of positive twist.
The vertical projection of the azimuthal component produces two
elongated polarities (``tongues'') which extend between the main ones.
The strength of these ``tongues'' is directly proportional to the
magnitude of the twist and their position depends on the sign of the
twist (the case with negative twist is a mirror image of the case with
positive twist).  The ``tongues'' are present only when the apex of
the flux tube is crossing the photosphere (during the period of ``flux
emergence'').  Later on, they disappear because the projection of the
azimuthal field in the vertical direction becomes less important.
Such picture was indeed present during the early evolution of AR 7912
(Fig.  \ref{f-pos-rot}.a and b).  The location of the ``tongues''
implies the presence of positive twist, in agreement with the
independent determination done previously with SXT and magnetic field
extrapolations.  The retraction of the ``tongues'' with time is
naturally explained by the emergence of the flux tube.  This implies a
positive rotation (counter-clockwise) of the mean position of the
positive polarity with respect to the negative one from October 13 to
18 (Fig.  \ref{f-rot}).

  The following negative (clockwise) rotation shows that the shape of
the emerging flux-tube is not planar, though for the flux-tube portion
which emerges during the few days of positive rotation the deviation
from planarity is small; thus, when describing above the ``tongues'',
we have neglected the influence of non-planarity.  It is only on the
long-term (three months, so an order of magnitude longer in time) that
the non-planarity of the flux tube becomes important.

\subsection{How was the flux tube formed~?}
\begin{figure*}[t]
\centering
\hspace{0.cm}
\includegraphics[bb=85 340 515 585,width=16.cm]{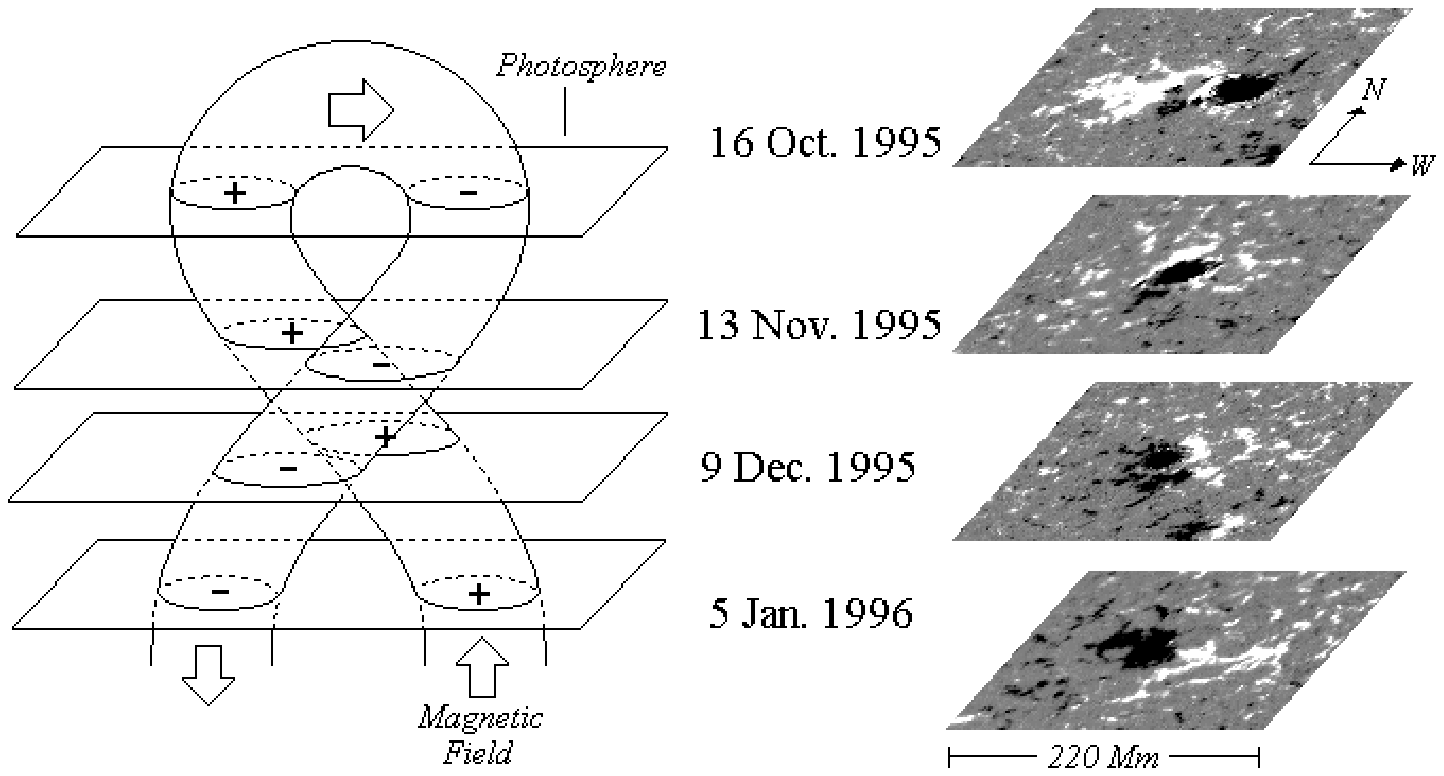}
       \caption{Evolution of AR 7912 during four rotations.  On the right,
photospheric longitudinal magnetograms are shown for each rotation
close to central meridian passage (CMP), positive (negative) values of
the field which appear in white (black) are saturated above (below)
50 G (-50 G).  The four frames have the same size ($\approx 220$ Mm).
On the left, a sketch of the magnetic flux-tube as deduced from the
observations.  The cuts by four horizontal planes show the approximate
location of the photosphere at the time of the magnetograms.}
\label{f-evolution}
\end{figure*}

  The non-classical evolution of AR 7912 has called our attention as a
probable candidate for a kink instability, since the flux-tube shape
deduced in Fig.  \ref{f-evolution} is similar to what is expected from
the non-linear development of this instability (Fan \etal 1999, Linton
\etal 1998, 1999).  However, the kink instability mode has certain
properties that may be verified in the observations.  In particular,
the handedness of the magnetic twist ($T$) and the writhe ($W$) of the
tube axis should be the same (e.g.  Fisher et al.  1999).  The twist
is the measure of the rotation of the field lines around the flux-tube
axis and the writhe is the measure of the rotation of the flux-tube
axis in space.  In a kink unstable flux-tube the twist ($T$) and the
writhe ($W$) should have the same sign because part of the twist is
transferred into the writhe when the kink instability develops.  For
AR 7912, the three month evolution implies a flux-tube axis with a
deformed helical shape (Fig.  \ref{f-evolution}).  The global negative
rotation gives a negative writhe for the flux-tube axis (it is on a
deformed left-handed helix).  The sign of the writhe is opposite to
the sign of the twist, thus a kink instability cannot be the origin of
the non-Hale nature of this region~!

  Can photospheric or shallow sub-photospheric large-scale flows be
the cause of the peculiar rotation of AR 7912~?  These motions may be
a photospheric vortex (like an earth tornado) with a negative
rotation, or they may result from the faster displacement of the
positive polarity around the negative one (e.g.  driven by the
magnetic tension of the flux tube).  However, this ``surface'' flow
cannot explain why the AR was initially formed with a non-Hale
orientation.  Moreover, the dispersion of the positive polarity (which
should be the leading in the southern hemisphere during solar cycle
22) is observed to be much faster than the dispersion of the negative
polarity along these four rotations, as can be seen in Fig.
\ref{f-evolution} (right portion).  This behavior is opposite to that
of Hale regions.  To quantify this we determined a flux-weighted mean
size of both polarities, $R_{p}$, $R_{n}$, with $R_{p}$ defined by:
\be \label{R}
  R_p = {\sum_{B_z > B_{min}} \sqrt{(x-X_{p})^{2} + (y-Y_{p})^{2} } B_z
          \over \sum_{B_z > B_{min}} B_z} \, ,
\ee
and a similar expression for $R_{n}$ (see Eq.  \ref{Pos} for the
definition of $X_{p}$, $Y_{p}$ and $B_{min}$).  Table
\ref{tbl_evradius} shows the evolution of $R_{p}$ and $R_{n}$ for $|
B_z | > 10$ G (which we consider a value more representative of the
relevant magnetic flux of the AR) for the four rotations.  The values
correspond to the averages of three days around CMP.  We want to
remark that the new small bipole appearing at the South of the region
was not included in the computations.

  Notice that the results of Table \ref{tbl_evradius} are opposite to
what is expected in Hale regions. In such normal regions, the longer
coherence of the leading polarity is explained by the evolution of the
$\Omega$-loop through the convective zone: the Coriolis force in an
ascending flux-tube pushes the plasma away from the preceding polarity
towards the following one; the resulting decrease of the plasma
pressure makes the leading spot more confined and with a stronger
magnetic field (Fan et al.  1993).  In AR 7912 we have observed just
the opposite evolution because the positive polarity was indeed in a
following position during the emergence (see Fig.  \ref{f-evolution})!
This faster dispersion of which should be the normal preceding
polarity shows that indeed the flux tube traveled through a
significant part of the convective zone (as classical $\Omega$-loops
do) in the reversed configuration, thus the peculiar rotation observed
is not due to a vortex motion at photospheric level (or just below).
Nevertheless, we cannot exclude the possibility that the rotation of
the AR is due to the action of the magnetic tension force restoring
the $\Omega$ loop shape (without the need of the continuous emergence
of the flux tube).

\placetable{tbl_evradius}

\begin{figure*}[t]
\centering
\hspace{0.cm}
\includegraphics[bb=90 300 480 550,width=16.cm]{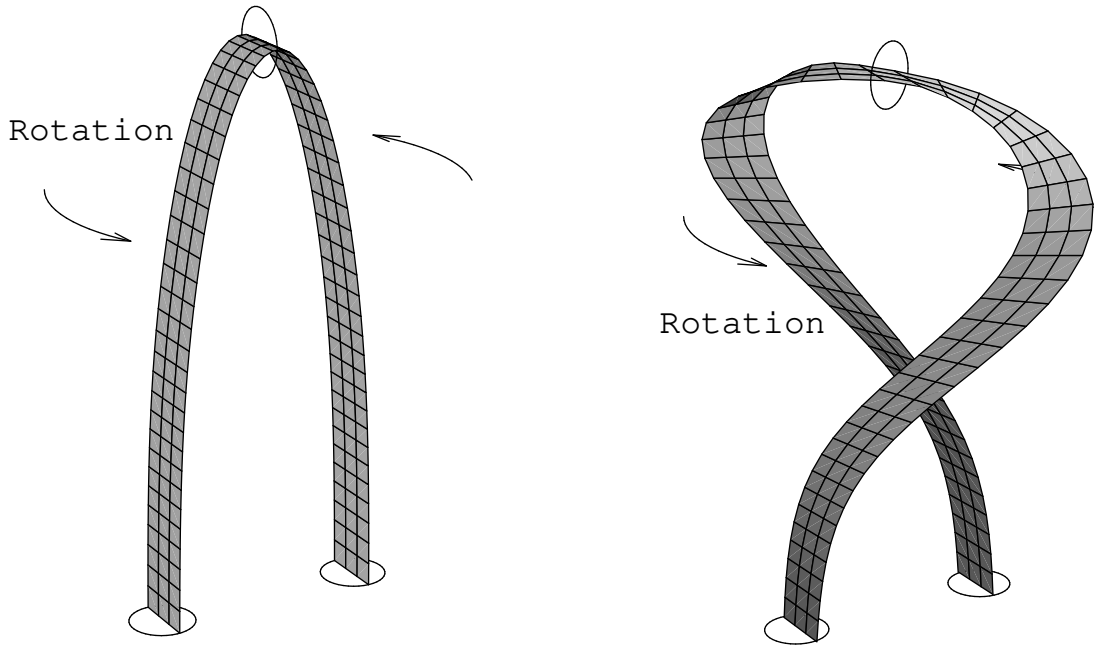}
       \caption{Sketch for the evolution of a buoyant flux-tube in the convective
zone.  (a) Shape of the flux tube after the Parker instability grows,
but before the effect of rotational motions.  Only a ribbon of field
lines is drawn (circles outline the shape of the tube) and the flux
tube is represented with no twist to simplify the drawing.  (b)
Evolution of the flux tube in the convective zone with no twisting
motions at its ``ends'', but with a deformation of its axis by
external rotational motions, giving a negative writhe to the flux tube
axis.  This induces a positive twist inside the flux tube.}
\label{f-cz}
\end{figure*}

   We propose that the origin of the peculiar evolution of AR 7912 may be
a simple interaction with convective motions.  As for other ARs, we
suppose that a flux tube forms in the convective overshoot region and
begins to rise as a normal $\Omega$-loop.  During its way through the
convective zone (including the bottom of the convective zone, but not
its top) the flux-tube axis is deformed by external motions which have
a rotational component (e.g.  a cyclonic flow or a strongly sheared
flow due to, e.g., an important local gradient of the differential
rotation).  From present data we cannot precise the delay between the
initial development of the Parker instability and the deformation of
the flux tube.  This delay can range from zero to significantly less
than the crossing time of the convective zone by the buoyant
flux-tube.  In any case, the motions are supposed to deform a finite
section of the axis into a helical shape with a negative handedness,
so a negative writhe ($\delta W$).  The conservation of magnetic
helicity implies that a positive twist ($\delta T = -\delta W$) is
induced in the flux tube (Moffatt and Ricca 1992, Berger and Field
1984).  For simplicity, the case with no initial $W$ and $T$ is shown
in Fig.  \ref{f-cz}.  In this picture, the only peculiarity of AR
7912, compared to other ARs, is that it is formed by a rising
flux-tube which encountered in its way convective motions which have a
rotational component.  All the peculiarities of AR 7912 (initial
non-Hale configuration, a positive rotation followed by a negative
one, a faster dispersion of what should be the leading polarity, a
non-monotonic variation of the distance between the polarities, an
opposite sign for $W$ and $T$) are the consequences of this
interaction.

  Earlier, Longcope et al. (1998) developed a model describing the creation
of magnetic twist in a flux tube from its interaction with helical
turbulence.  The convective flows are coupled to the flux tube by the
drag force and they progressively introduce an helicoidal deformation
in the flux-tube axis (writhe).  By conservation of magnetic helicity
a twist of opposite magnitude is introduced in the flux tube.  Such
effect, named the $\Sigma$ effect, predicts a hemispherical rule, a
magnitude as well as a statistical dispersion of the magnetic twist
similar to what is observed.  We suggest that the evolution of AR 7912
follows the model developed by Longcope et al.  (1998).

   We have given above a simple explanation of the behavior of AR 7912
supposing implicitly no initial $T$ and $W$ (Fig.  \ref{f-cz}.a) in
order to explain the main effects.  We make this view more precise
below.  The observations show that the flux tubes forming most active
regions have both significant magnetic writhe and twist at the
photospheric level (e.g. Canfield and Pevtsov, 1998).  The writhe is
thought to come from the effect of the Coriolis force (leading to
Joy's law) and from the interaction with the helical turbulence on the
ascending flux tube (Longcope et al.  1998).  The sub-photospheric
origin of the twist is thought to be at the core-convection zone
interface (e.g.  Gilman and Charbonneau, 1999).  An initial twist is
indeed needed so that the buoyant flux-tube is not destroyed by the
hydrodynamic vortex, which develops behind it during its transit
through the convective zone (Emonet \& Moreno-Insertis, 1998; Fan
\etal 1998).

 The conservation of helicity then simply writes:
 \begin{equation} \label{T+W}
 T_t+W_t = T_0+W_0  = H_0 \,
 \end{equation}
where $T_t$ and $W_t$ (resp.  $T_0$ and $W_0$) are the twist and the
writhe at time $t$ (resp.  initial time).  Taking a positive initial
helicity $H_0$ to study a typical flux tube in the southern hemisphere
(the negative case is symmetric), we have the following
possibilities:\\
  \indent - {\bf Case a}: $W_t < 0$, so the writhe created by the convective
motions
is negative and the twist of the emerging flux tube is positive
($W_t T_t<0$  and $T_t>H_0>0$),\\
  \indent - {\bf Case b}: $0< W_t < H_0$, so the added writhe is not enough to
create a negative twist and both are of the same sign
($W_t T_t>0$  and $H_0>T_t>0$),\\
  \indent - {\bf Case c}: $W_t > H_0$, so the writhe created by the convective
motions is large enough and positive to create a negative twist
($W_t T_t<0$,  $H_0>0$ and $T_t<0$). \\
   The observable photospheric evolution of the longitudinal
magnetic field resulting from the emergence of the twisted flux-tube is
determined by the twist ($T_t$) and the writhe ($W_t$) of the flux
tube. As the apex of the twisted flux tube emerges, two elongated
magnetic polarities (``tongues'') appear (see Sect. 3.2). This
phenomenon, which is determined by the twist of the flux tube, 
gives a relative rotation of the mean position of the photospheric magnetic 
polarities and
lasts approximately as long as the period of ``flux emergence'' (several
days) (see D\'emoulin et al. 2000). Once the apex of the flux tube has 
fully emerged, the tongues disappear, and as the flux tube continues 
emerging, the flux tube shape, so
its writhe, determines the long-term evolution of the magnetic
polarities (Sect. 3.1). Then we have:\\
  \indent - {\bf Case a}: an initial short-term positive rotation of the mean
positions of the polarities followed by a long-term negative rotation 
(case of AR 7912),\\
  \indent - {\bf Case b}: a continuous positive rotation,\\
  \indent - {\bf Case c}: an initial short-term negative rotation followed 
by a long-term positive rotation. \\
   In all these cases, the long-term rotation brings back the polarities to the
Hale orientation. Case c would appear as an exception to the
hemispherical rule for the sign of $\alpha$ (while the flux tube
is initially formed in the same way at the bottom of the convective zone).
Finally, in Case b, the origin of the peculiar geometry of the tube
can be confused with a kink instability.

\section{Conclusion}
\label{conclusion}

   We have shown that AR 7921, AR 7930 and the bipolar field on January
1996, at $\approx$ S10, are in fact the re-appearances on the solar
disk of AR 7912.  Emphasizing this fact, we use the same NOAA number
(AR 7912) for all four regions.

   We have focused this study on the long-term (4 rotations) evolution of
AR 7912 because it is a non-Hale region, which showed unusual
rotation of the positive polarity with respect to the negative one,
making approximately half a turn.  Apart from that, AR 7912 is a
simple active region from the point of view of the magnetic
complexity, since it is basically a bipolar configuration.  Thus, AR
7912 gives us the possibility to analyze the emergence of one simple
magnetic flux-tube which undergoes unusual motions.  From the
photospheric evolution of AR 7912, we deduce that the flux-tube axis
has a helical shape.
A possible interpretation of this evolution may be a non-linear
development of the kink instability, although further analysis does
not support that, since the writhe of the flux tube and its internal
twist do not have the same sign (as they should have for a kink
instability mode). Next we exclude the photospheric (or shallow
sub-photospheric) vortex motions, because faster dispersion of
positive polarity implies that the flux tube has traveled through
at least part of the convection zone in such an helical shape
(Sec. 3.3).
The most likely origin of the peculiar flux-tube geometry of AR 7912
is an interaction, deep down in the convective zone, of the rising
flux-tube with plasma motions having a rotational component.  Such
motions would bring the flux-tube axis in a left-handed helix-like
shape, giving a negative writhe to the axis.  This writhe is observed
as a negative (clockwise) rotation of the photospheric polarities as
the flux tube emerges.  The conservation of magnetic helicity induces
a positive twist which adds up to the initial twist (probably, also
positive due to the hemispheric rule for the helicity).  With the
hypothesis of the deformation of the flux-tube axis by convective
motions, all the peculiarities of AR 7912 (initial non-Hale
configuration, a positive rotation followed by a negative one which
brings back the region to the Hale orientation, a faster dispersion of
which should be the leading polarity, a non-monotonic variation of the
distance between the polarities, an opposite sign of writhe and twist)
are explained in a logical way.

\begin{figure*}[t]
\centering
\hspace{0.cm}
\includegraphics[bb=90 300 490 510,width=15.cm]{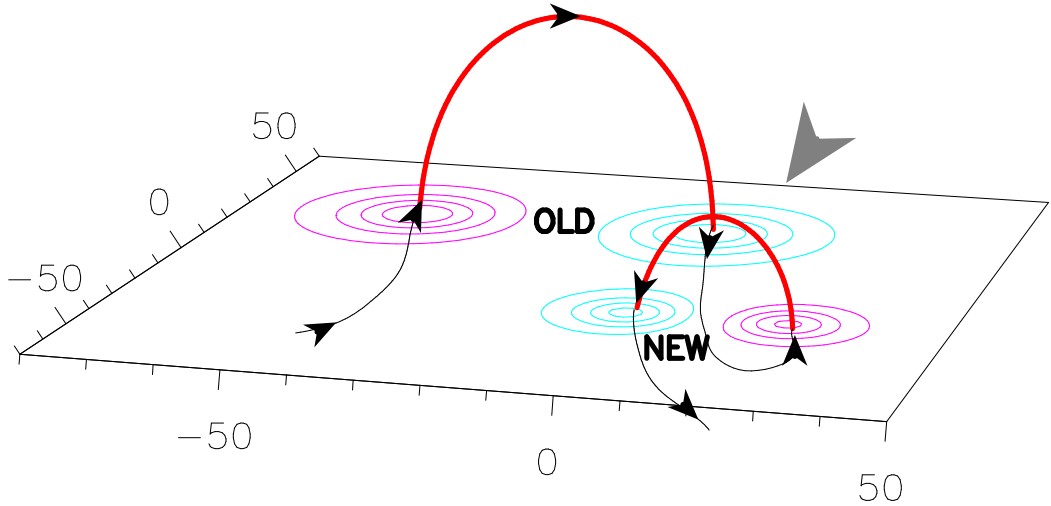}
       \caption{Possible magnetic configuration for the two cases analyzed by Pevtsov
\& Longcope (1998).  Their first case is represented with the labels
``old'' for AR 7918 and ``new'' for AR 7926 to indicate which active
region emerged first (see their Fig.  2).  It is enough to inverse the
position of ``new'' and ``old'' and the sign of the polarities, since
they are on different hemispheres, to get the basic configuration of
their second example (i.e.  ``old'' for AR 7091 and ``new'' for AR
7123, see their Fig.  9).  We suggest that this configuration was
formed in the same way we propose in Fig.  6 for the region called
``new'' or, alternatively, by a deformed $\Omega$-loop pushed down
close its central portion as indicated by the largest grey arrow (see
text). The portion of the flux-tube axis above (below) the photosphere
is shown with a thick (thin) continuous line. Dark (light) grey
isocontours correspond to the positive (negative) polarities.}
\label{f-pevtsov}
\end{figure*}

   We bring now the case of AR 7912 in a broader context to analyze
whether such evolution has been reported in other cases.  We
concentrate now, in particular, on the possible origin of these
peculiarities: kink instability or interaction with convective motions
(photospheric motions are unlikely to be the origin as discussed
above).  Pevtsov \& Longcope (1998) show two examples where a Hale AR
is associated with a non-Hale AR. The latter appears in the
consecutive solar rotation.  In Fig.  \ref{f-pevtsov} we show a 3-D
perspective of the scenario proposed by them.  These authors do not
explicitly address the mechanism which created this kinked flux tube,
but an a priori possibility may be the development of the kink
instability (e.g.  Fan \etal 1999, Linton \etal 1999, Matsumoto \etal
1998).  However, the observational results of Pevtsov and Longcope
rather show that the magnetic flux tube has not an helical shape, but
is simply a classical $\Omega$-loop bended down close to its center
(see Fig.  \ref{f-pevtsov}).  The magnetic linkage shown in Fig.
\ref{f-pevtsov} is deduced from the degree of dispersion of the
polarities, but we notice that our conclusion would be the same if we
exchange the linkage of each positive polarity to the other
alternative negative polarity.  We suggest that the origin of this
configuration may be the interaction of the rising $\Omega$-loop with
plasma motions in the convective zone, a similar scenario to the case
presented by us.  A first possibility is that the Parker instability
developed successively at two nearby portions of the same toroidal
flux-tube, and that the non-Hale region was formed by the same
mechanism as for the AR studied in this paper.  A second possibility
is that one rising $\Omega$-loop interacted with downward convective
motions along its central portion.  One side of the loop was pushed
down more efficiently than the other one leading to a shift in the
emerging time (of about one solar rotation) between the two sides (so,
the two linked ARs).  Both examples that Pevtsov and Longcope provide
can be interpreted by one of the above possible mechanisms.  The only
minor differences between their two examples are: first, that the sign
of the polarities is opposite because the ARs are located in different
hemispheres, and, second, that the southern part of the $\Omega$-loop
is emerging less rapidly than the northern part in the first example
(AR 7918 and AR 7926), while the reverse is true in the second example
(see Fig.  \ref{f-pevtsov}).

  Other case studies have been carried out on much shorter time scales
(few days compared to the four solar rotations described here).
Tanaka (1991) presented two cases of peculiar active-region evolution.
The first one (July 1974) is characterized by showing both directions
of rotation (clockwise and counter-clockwise), it is a complex case
which cannot be easily explained by the kink instability mode.  The
second case (August 1972) presents a negative (clockwise) rotation of
the main bipole, so a writhe of the same sign as the twist (as deduced
from the shape of the H$\alpha$ fibrils and flare ribbons).  This is a
good candidate for the kink instability mode.  Other studies have been
rather focused on the short-term evolution (few days) of small bipoles
within an active region.  Evidences for the emergence of twisted flux
tubes have been found in several active regions (Kurokawa, 1987; Ishii
et al.  1998).  Lites \etal (1995) have interpreted their observations
in terms of an ascending closed ball of twisted field.  From these
papers it is difficult to draw any conclusion on the writhe and twist.
Leka \etal (1996) found at least four small bipoles which emerged
twisted and have the same sign for the writhe and the twist, thus
being good candidates for the kink instability.

 Canfield and Pevtsov (1998) have investigated the statistical
relationship between the twist and the writhe of 91 ARs following the
Hale law (8 non-Hale ARs of the sample have been eliminated from the
statistics).  The twist is determined by the best fit of the vector
magnetic field data with a linear force-free extrapolation, and the
writhe is deduced from the tilt of the active region axis with respect
to the solar equator.  In the south hemisphere they found that the
majority of the ARs have a negative writhe (Joy's law) and a positive
twist based on independent statistics (note that the sign defined from
the writhe is opposite by definition to the sign of the tilt angle).
But surprisingly, the correlation between writhe and twist was found
to be direct, so that an AR with positive twist has more chances to
have a positive writhe than a negative one; the opposite of what is
expected from the two independent statistics in function of latitude
mentioned above.  This is in favor of the above Case b (where the
added writhe is not sufficient to reverse the sign of the initial
twist) or of the kink instability (where the writhe comes from the
initial twist).  However, the statistical laws are weak (a large
dispersion is present); a fact which may explain their apparent
incompatibility.  Such large dispersion probably comes from the
interaction between ascending flux tubes and turbulent motions
(Longcope et al.  1998).

  Looking at all the cases listed above, the basic characteristic of
the kink instability ($WT>0$) is only present in some examples; which
can also be interpreted as examples of our Case b (where the writhe is
not sufficient to reverse the sign of the initial magnetic helicity).
Moreover, there are cases which do not have the characteristics of the
kink instability.  Our studied case (AR 7912) has opposite writhe and
twist, and the two examples of Pevtsov and Longcope (1998) are likely
to be explained also by an interaction of the rising flux-tube with
convective motions.  With so few examples studied in detail, we
certainly cannot confirm that all cases have a common origin, that is
the interaction of a normal rising $\Omega$-loop with convective
motions, even if such interpretation is attractive (see also Longcope
et al.  1998).  Clearly, an extension of the study of peculiar cases
(non-Hale active regions) is needed.  The cases which show the same
sign for the twist and the writhe need a further analysis to confirm
or not that they are indeed caused by the development of the kink
instability.  Then, following Pevtsov and Canfield (1998), a further
statistical analysis of a large sample of ARs is needed.

   Finally, we note that if it would be shown that the kink
instability is not the origin of the formation of non-Hale ARs, this
will have a positive outcome in helping to determine more precisely
the twist in buoyant flux-tubes at the base of the convective zone.
The threshold to reach the kink instability decreases like the inverse
of the flux-tube radius (e.g.  Linton \etal 1998); then, the kink
instability is easier to achieve as the flux tube moves upward and
expands (by at least a factor of 10).  So, in order to prevent the kink
instability to develop during the ascent in the convective zone, the
initial twist should be low enough.  On the other hand, the flux tube
needs to be slightly twisted initially to keep its coherence through
the convective zone (Emonet and Moreno-Insertis 1998 and Fan \etal
1998).  Thus, if the kink instability does not develop, it is probable
that the initial twist of flux tubes emerging at the photosphere lies
in a narrow range, just above the minimum twist needed to maintain the
coherence of the tube.  To test precisely this conclusion, present
numerical simulations need to be extended to follow a flux tube
through many gravitational scale-heights (from the bottom of the
convective zone to the photosphere).  Then, even if careful
examination of the observations would show that the kink instability
is less important than it is presently thought, further investigations
in this area are clearly needed.

\acknowledgments
M.L.F, C.H.M. and P.D. acknowledge financial support
from ECOS (France) and ANPCYT (Argentina) through their cooperative
science program (A97U01).  LvDG acknowledges the Hungarian Government
Grants TP 096 OTKA T026165, T032846 and AKP 97-58 2,2.  We thank the
anonymous referee for the constructive and helpful comments which
improved our paper.  The NSO/Kitt Peak data used here are produced
cooperatively by NSF/NAO, NASA/GSFC, and NOAA/SEL. The {\it Yohkoh}
Soft X-ray Telescope is a collaborative project of the Lockheed Palo
Alto Research Laboratory, the National Astronomical Observatory of
Japan, and the University of Tokyo, supported by NASA and ISAS.

\clearpage
\begin{deluxetable}{lcc}
\tablecaption{Evolution of the mean size of the polarities.
\label{tbl_evradius}}
\tablehead{
\colhead{Rotation Number} & \colhead{$R_p$ (Mm)} & \colhead{$R_n$ (Mm)}}
\startdata
  1st. & 37.  & 24. \nl
  2nd. & 48.  & 24. \nl
  3rd. & 58.  & 34. \nl
  4th. & 62.  & 51. \nl
\enddata
\end{deluxetable}

\clearpage

\clearpage

\clearpage

\clearpage

\clearpage

\clearpage

\clearpage

\end{document}